\documentclass{aa}
\usepackage{color}
 \newcommand{\Htwo}{$\mathrm{H}_2$}
 \newcommand{\HI}{\ion{H}{i}}
 \newcommand{\COzero}{${}^{12}\mathrm{CO}(1-0)$}

 \newcommand{\Ox}{[\ion{O}{iii}]}

\def\lae{\mathrel{<\kern-1.0em\lower0.9ex\hbox{$\sim$}}}
\def\gae{\mathrel{>\kern-1.0em\lower0.9ex\hbox{$\sim$}}}

\usepackage{graphicx}
\usepackage{txfonts}

\begin{document}

   \title{The Close AGN Reference Survey (CARS)}

   \subtitle{No evidence of galaxy-scale hot outflows in two nearby AGN\thanks{The scientific results reported in this article are based on observations made by the \textit{Chandra} X-ray Observatory (PI: G.~Tremblay, ID: 17700519), and on observations collected at the European Organization for Astronomical Research in the Southern Hemisphere under ESO program(s) 094.B-0345(A) (PI: B.~Husemann).}}

   \author{M.~C.~Powell \inst{1}
   \and B. Husemann \inst{2}
   \and G.~R.~Tremblay \inst{3}
   \and M. Krumpe \inst{4}
   \and T. Urrutia \inst{4}
   \and S.~A.~Baum \inst{5,6}
   \and G. Busch \inst{7}
   \and F. Combes \inst{8}
   \and S.~M.~Croom \inst{9}
   \and T.~A.~Davis \inst{10}
   \and A.~Eckart \inst{7,11}
   \and C.~P.~O'Dea \inst{5,12}
   \and M. Pérez-Torres \inst{13,14} 
   \and J. Scharw\"achter \inst{15}
   \and I.~Smirnova-Pinchukova\inst{2}
   \and C.~M.~Urry \inst{1}
          }

   \institute{Yale Center for Astronomy and Astrophysics, and Physics Department, Yale University, PO Box 2018120, New Haven, CT 06520-8120, USA\\
    \email{meredith.powell@yale.edu}
    \and
 Max-Planck-Institut für Astronomie, K\"onigstuhl 17, D-69117 Heidelberg, Germany
    \and 
      Harvard-Smithsonian Center for Astrophysics, 60 Garden St., Cambridge, MA 02138, USA
      \and 
      Leibniz-Institut f\"ur Astrophysik Potsdam, An der Sternwarte 16, 14482 Potsdam, Germany
      \and
      Department of Physics \& Astronomy, University of Manitoba, Winnipeg, MB R3T 2N2, Canada
      \and
      Carlson Center for Imaging Science,  Rochester Institute of Technology, Rochester, NY 14623 USA
      \and
      I. Physikalisches Institut der Universität zu Köln, Zülpicher Str. 77, 50937 Köln, Germany
      \and
      LERMA, Observatoire de Paris, College de France, PSL Univ., CNRS, Sorbonne Univ.,  75014 Paris, France
      \and 
      Sydney Institute for Astronomy, School of Physics, A28, The University of Sydney, NSW, 2006, Australia
       \and 
      School of Physics \& Astronomy, Cardiff University, Queens Buildings, The Parade, Cardiff, CF24 3AA, UK
      \and
      Max-Planck-Institut für Radioastronomie, Auf dem Hügel 69, 53121, Bonn, Germany
      \and 
      School of Physics \& Astronomy, Rochester Institute of Technology, Rochester, NY 14623 USA
      \and
      Instituto de Astrofísica de Andaluc\'{i}a, Glorieta de las Astronom\'{i}a s/n, 18008 Granada, Spain
      \and 
      Departamento de F\'{\i}sica Te\'orica, Facultad de Ciencias, Universidad de Zaragoza, E-50009 Zaragoza, Spain
      \and 
      Gemini Observatory, Northern Operations Center, 670 N. A’ohoku Pl., Hilo, Hawaii, 96720, USA
             }

 \date{Received 14 May 2018 / Accepted 30 June 2018}
 
  \abstract
   {}  
   {
  We probe the radiatively-efficient, hot wind feedback mode in two nearby luminous unobscured (type 1) AGN from the Close AGN Reference Survey (CARS), which show intriguing kpc-scale arc-like features of extended \Ox\ ionized gas as mapped with VLT-MUSE.  We aimed to detect hot gas bubbles that would indicate the existence of powerful, galaxy-scale outflows in our targets, HE~0227-0913 and HE~0351+0240, from deep (200 ks) {\it Chandra} observations.
   }
   {By measuring the spatial and spectral properties of the extended X-ray emission and comparing with the sub kpc-scale IFU data, we are able to constrain feedback scenarios and directly test if the ionized gas is due to a shocked wind.}
   {No extended hot gas emission on kpc-scales was detected. Unless the ambient medium density is low ($n_{H}\sim~1$ cm$^{-3}$ at 100 pc), the inferred upper limits on the extended X-ray luminosities are well below what is expected from theoretical models at matching AGN luminosities.}
   {We conclude that the highly-ionized gas structures on kpc scales are not inflated by a hot outflow in either target, and instead are likely caused by photoionization of pre-existing gas streams of different origins. Our nondetections suggest that extended X-ray emission from an AGN-driven wind is not universal, and may lead to conflicts with current theoretical predictions.}

   \maketitle
%
%_________________________________________________

\section{Introduction}

Supermassive black holes undergo phases of extreme accretion, observed as active galactic nuclei (AGN), which give off enormous amounts energy that is deposited into the host galaxy. Tight correlations between black hole and galaxy properties suggest a coevolution between them, in which the energy couples to the surrounding environment and may influence the state of star formation in the galaxy \citep{Silk:1998}. Such feedback is generally invoked in simulations to recover the observed local luminosity function, galaxy bimodality in color space, and the number density of massive galaxies \citep[e.g.,][]{DiMatteo:2005,Bower:2006,Schaye:2015,Weinberger:2017}.

One manifestation of AGN feedback is the radiatively-efficient hot wind mode, which has been predicted in semi-analytical \citep[e.g.,][]{Weinberger:2017} and analytical models \citep{FQ:2012,Zubovas:2012}.
This is when radiation pressure from the accretion disk pushes the surrounding gas outward, driving a hot, shocked, bipolar wind if overpressured. Such a galactic `superwind' would sweep up and shock the surrounding ambient gas as it expands, possibly causing a suppression of star formation if energetic enough to expel the gas from the galaxy. Observationally, this outflow would be characterized by thin ionized shells surrounding a hot X-ray-bright bubble (Heckman 1990), similar to the galaxy-scale winds driven by supernovae and/or massive stars. Analytical models of AGN-driven winds have been proposed \citep{FQ:2012} based on observations of nearby quasars, with observational signatures that can distinguish them from stellar winds. Specifically, the luminosity and morphology of spatially-extended X-ray emission from the forward shock into the surrounding interstellar medium (ISM) is predictable, given assumptions on the surrounding ambient density profile \citep{Nims:2015} and the coupling efficiency between the AGN and its environment.

\begin{figure*}
 \sidecaption
 \includegraphics[width=\textwidth]{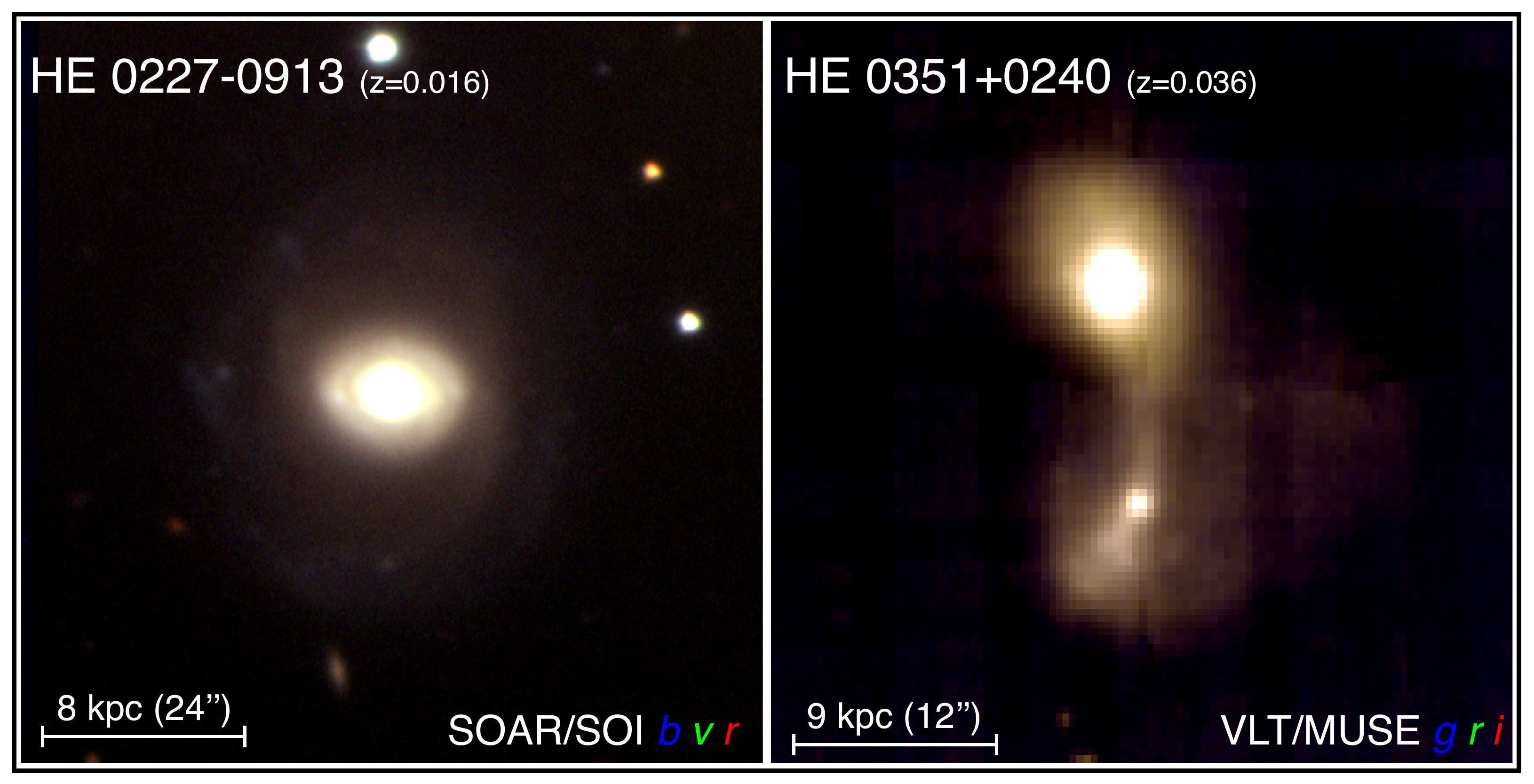}
 \caption{Three-color composite images of the two galaxies for which we present new \textit{Chandra} and MUSE data in this paper. {\it Left:} SOAR Optical Imager (SOI) $b$-, $v$-, and $r$-band composite of  HE 0227-0913 (Markarian 1044). These bands have been assigned to the blue, green, and red channels, respectively. {\it Right:} Composite made by collapsing three slices through the MUSE cube across wavelength ranges that roughly correspond to the $g$-, $r$-, and $i$-bands, again respectively assigned to the blue, red, and green channels. We have applied a logarithmic stretch to both panels to better show the low surface brightness galaxy outskirts alongside the bright type 1 nuclei. Both images have been rotated, with east left and north up. }
 \label{fig:MUSE_color}
\end{figure*}

There have been previous observations of extended hot X-ray emission around nearby AGN, for which radiative feedback is consistent. This includes the luminous, obscured radio-quiet quasar SDSS J135646.10+102609.0 \citep{Greene:2014}, selected for its high intrinsic luminosity and strong and extended ionized gas. Additionally there is the Type 2 `Teacup AGN' (SDSS J1430+1339; \citealt{Lansbury:2018}),
also selected for its luminous, extended, as well as broad (FWHM$>700$ km/s) \Ox\ emission, along with the nearest ultraluminous infrared galaxy (ULIRG) Mrk 231 \citep{Veilleux:2014}. The fading obscured AGN IC 2497 (`Hanny’s Voorwerp' system; \citealt{Sartori:2016}) also shows evidence for hot, diffuse gas from a past high accretion state. However the ubiquity of quasar-driven superbubbles is currently unclear, especially for unobscured AGN whose X-ray emission is dominated by the central engine. Deep, spatially resolved X-ray data is currently lacking for a representative sample of such objects.

In this paper we present a study of two luminous Type 1 AGN aiming to explore the connection of ionized gas filaments on kpc-scales and a presumed expanding hot gas bubble powered by the AGN radiation.  We combine 200 ks of new {\it Chandra} X-ray observations in conjunction with optical integral-field unit (IFU) observations, in order to distinguish between the hot gas outflow scenario and the case where the \Ox\ arcs are gas clouds illuminated by the AGN, which have been previously expelled from the galaxy. Throughout this paper we assume $H_0 = 70$ km s$^{-1}$ Mpc$^{-1}$, $\Omega_M = 0.3$, and $\Omega_{\Lambda} = 0.7$.

%__________________________________________________________________

\section{The Close AGN Reference Survey}
The {\bf C}lose {\bf A}GN {\bf R}eference {\bf S}urvey (CARS; \citealt{Husemann:2017}, Husemann et al. in prep.\footnote{\texttt{www.cars-survey.org}}) aims to constrain the galaxy-scale feeding vs. feedback in a larger sample of nearby luminous AGN via spatially-resolved observations across the entire X-rays to radio electromagnetic spectrum.  The CARS targets are drawn from a mother sample of 99 luminous AGN detected in the Hamburg-ESO survey \citep[HES,][]{Wisotzki:2000} up to $z<0.06$, of which 39 galaxies were targeted with the IRAM 30m telescope to characterize the molecular gas content via \COzero\ line observations \citep{Bertram:2007}. Focusing on the \citeauthor{Bertram:2007} sub-sample, CARS has undertaken a full 3D (spectral+spatial) mapping campaign in the optical, primarily using the Multi-Unit Spectroscopic Explorer \citep[MUSE,][]{Bacon:2010,Bacon:2014a} at the Very Large Telescope (VLT). Those observations are being complemented with additional radio, IR, optical, UV and X-ray observations to get a complete picture of the stellar and multiphase gas content of the host galaxies in relation to the AGN properties.  

From the CARS sample we selected two galaxies, HE~0227$-$0913 (also known as Mrk~1044) and HE~0351$+$0240, for deep pilot {\it Chandra} observations to detect extended hot diffuse gas. The high accretion rates of the AGN and detection of kpc-scale ionized gas filaments make both galaxies good candidates to test the presence of expanding X-ray bubbles as expected for the radiative AGN feedback mode. The two galaxies presented here are complementary in studying the SMBH-galaxy connection, as they have contrasting host galaxy properties and are in very different evolutionary stages. HE~0227$-$0913 is a narrow-line Seyfert 1 galaxy (NLS1) at $z=0.016$ with a disk-like morphology. By reverberation mapping, its black hole mass is estimated to be $1.4\times 10^6~M_{\odot}$ \citep{Wang:2014}. HE~0351$+$0240, on the other hand, is in an ongoing major merger at $z=0.036$, with an estimated black hole mass of $1.3\times 10^7~M_{\odot}$ \citep{Koss:2017}. The characteristics of these galaxies are summarized in Table \ref{table:1}.

\begin{table}      
\small
\caption{Basic parameters for each galaxy.} 
\label{table:1}      
\centering                                   
%\begin{tabular*}{.5\textwidth}{@{}c c c c c c@{}}    
\begin{tabular*}{.5\textwidth}{c c c c c c}          
\hline\hline                        
Galaxy & $z$ & $M_\star$ \tablefootmark{a} & $M_{\rm{BH}}$ \tablefootmark{b} & $i$\tablefootmark{c} & kpc/$\arcsec$ \\ 
 & &($M_{\odot}$) &($M_{\odot}$) & & \\
\hline\smallskip
 
 HE~0227$-$0913    & 0.016 & $1.3\times 10^{10}$ & $1.4\times 10^{6}$ & $67\pm5$ & 0.33\\\smallskip 

HE~0351$+$0240    &  0.036 & $3.2\times 10^{10}$ & $1.3\times 10^{7}$ & N/A & 0.72 \\ 
\hline                                       
\end{tabular*}
\tablefoot{
\tablefoottext{a}{Stellar masses based on the QSO-subtracted $g$-$i$ host color and $M_i$ absolute host magnitude following \citet{Taylor:2011}.}
\tablefoottext{b}{BH masses taken from \citet{Wang:2014} for HE~0227$-$0913 and \citet{Koss:2017} for HE~0351$+$0240.}
\tablefoottext{c}{Galaxy disk inclination estimated only for HE~0227$-$0913 from the gas kinematic velocity. }
}
\end{table}

\begin{figure*}
  \includegraphics[width=\textwidth]{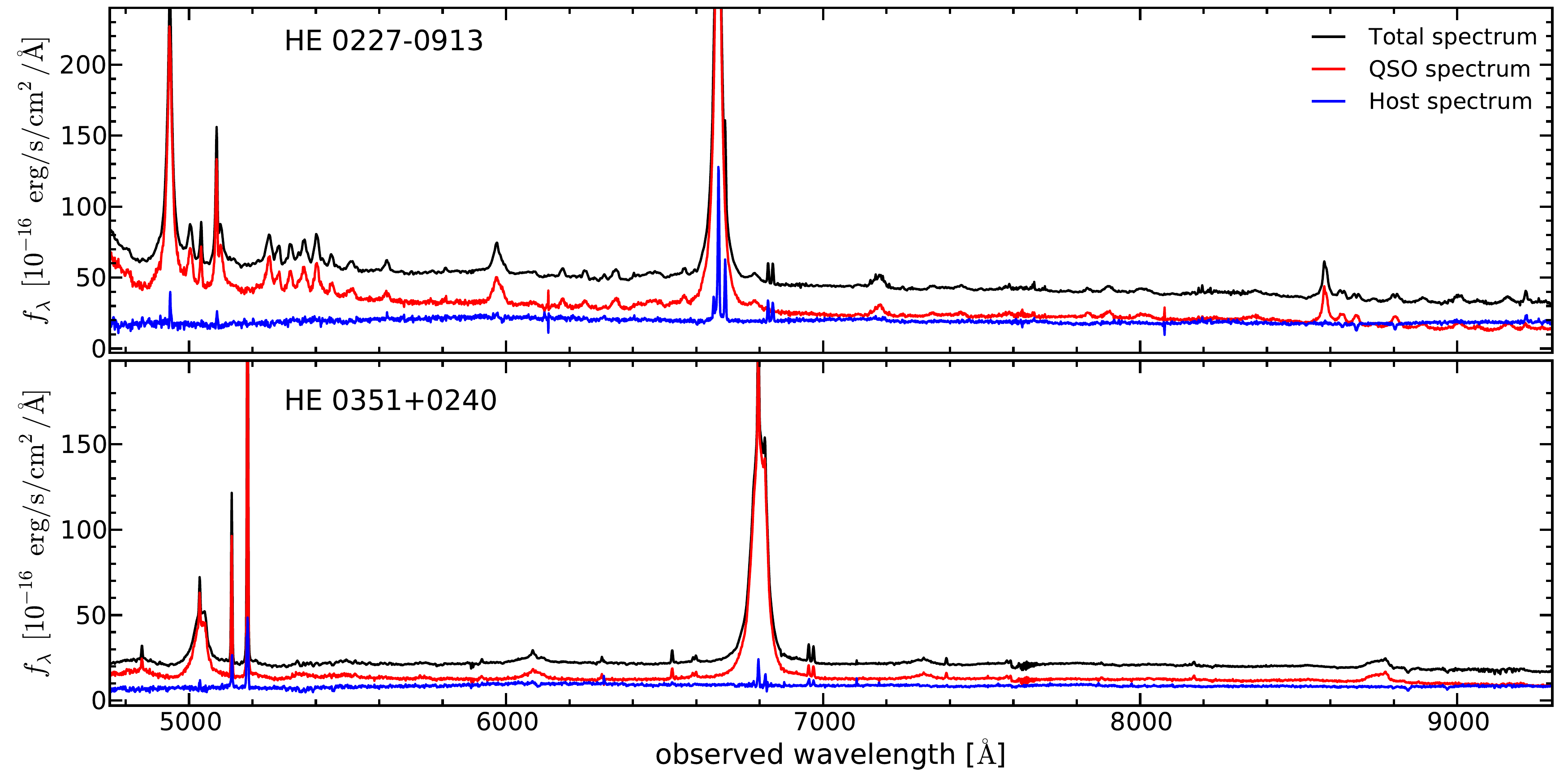}
 \caption{Integrated central MUSE aperture spectra within a diameter of $3\arcsec$ for HE~0227$-$0913 and HE~0351$+$0240. The total spectrum (black lines) is decomposed into a point-like QSO and extended host-galaxy contribution using the nonparametric QSO-host galaxy decomposition algorithm implemented in  QDeblend${}^\mathrm{3D}$ \citep{Husemann:2013,Husemann:2014}.}\label{fig:decomp}
\end{figure*}

\section{Observations and data reduction }
\subsection{MUSE integral-field spectroscopy}
We observed HE~0227$-$0913 and HE~0351$+$0240 with MUSE at the Very Large Telescope (VLT) the nights of 10 January 2015 and 26 December 2014, respectively. The two targets were observed for 1200\,s and 1600\,s integration times, each split into two independent exposures which were rotated by 90\degr\ with a small dither offset to aid in the cleaning of cosmic rays and flat-fielding artifacts. The wide-field mode (WFM) of MUSE provides a nearly rectangular field-of-view (FoV) with a size of $1\arcmin\times1\arcmin$ and a sampling of $0\farcs2$. The wavelength range is $4750\AA$--$9300\AA$ with a spectral resolution of $\sim$2.5\AA\ that slightly varies with wavelength \citep{Bacon:2017,Guerou:2017}. 

The MUSE data were reduced with the official ESO pipeline \citep[Version 1.6,][]{Weilbacher:2012} performing standard calibration tasks, such as bias subtraction, slice extraction, flat-fielding, wavelength calibration using arc lines, and flux calibration based on spectrophotometric standard stars. The fully calibrated individual exposures were then combined and resampled using the `drizzle' algorithm \citep{Fruchter:2002} to a 3D datacube with rectified wavelength range. Sky subtraction was done by creating a master sky spectrum from blank-sky regions within the MUSE FoV. We further suppressed sky-line residuals with a custom-made principal component analysis approach like ZAP \citep{Soto:2016}, but optimized and simplified for the given purpose and data set (Husemann et al. in prep.). 

Figure~\ref{fig:MUSE_color} shows the continuum light for HE~0227$-$0913 and HE~0351$+$0240. The $B$, $V$, and $R$ three-color composite of HE~0227-0913 is shown, taken as part of an optical and NIR followup campaign for the CARS galaxies with the SOAR Optical Imager (SOI; \citealt{Walker:2003}) on the SOAR telescope, (SOAR/NOAO program 2016A-0006 PI: G.~Tremblay).
For HE~0351+0240, the color image was created from the $gri$ broadband image extracted from the MUSE datacube. HE~0227$-$0913 appears as an inclined disk galaxy with a rather faint extended disk and spiral arms, while HE~0351$+$0240 appears as an ongoing major merger system where the two galaxy cores are still separated.

\subsection{\emph{Chandra} X-ray spectroscopy}
We observed HE~0227$-$0913 with the \emph{Chandra} Advanced CCD Imaging Spectrometer (ACIS-S; PI: G. Tremblay) on 12 October 2015 (obs. ID 18188), 10-14-2015 (obs. ID 18684), and 16 October 2015 (obs. ID 18685) for a total of 93.4 ks. HE~0351$+$0240 was observed on 14 December 2015 (obs. ID 18189) and 15 December 2015 (obs. ID 18726) for a total of 93.7 ks. Due to the high flux of the objects, we used a 1/8 subarray to minimize pileup. The data were reduced and analyzed with the CIAO 4.9 software \citep{CIAO:2006}. The `{\tt chandra\_repro}' script was used with the CALDB 4.7.3 set of calibration files to reprocess the level 1 event files. We filtered the data to eliminate cosmic rays and bad grades, and verified no significant changes in the background with time.

\section{Data analysis}

\begin{figure*}[h]
 \includegraphics[width=\textwidth]{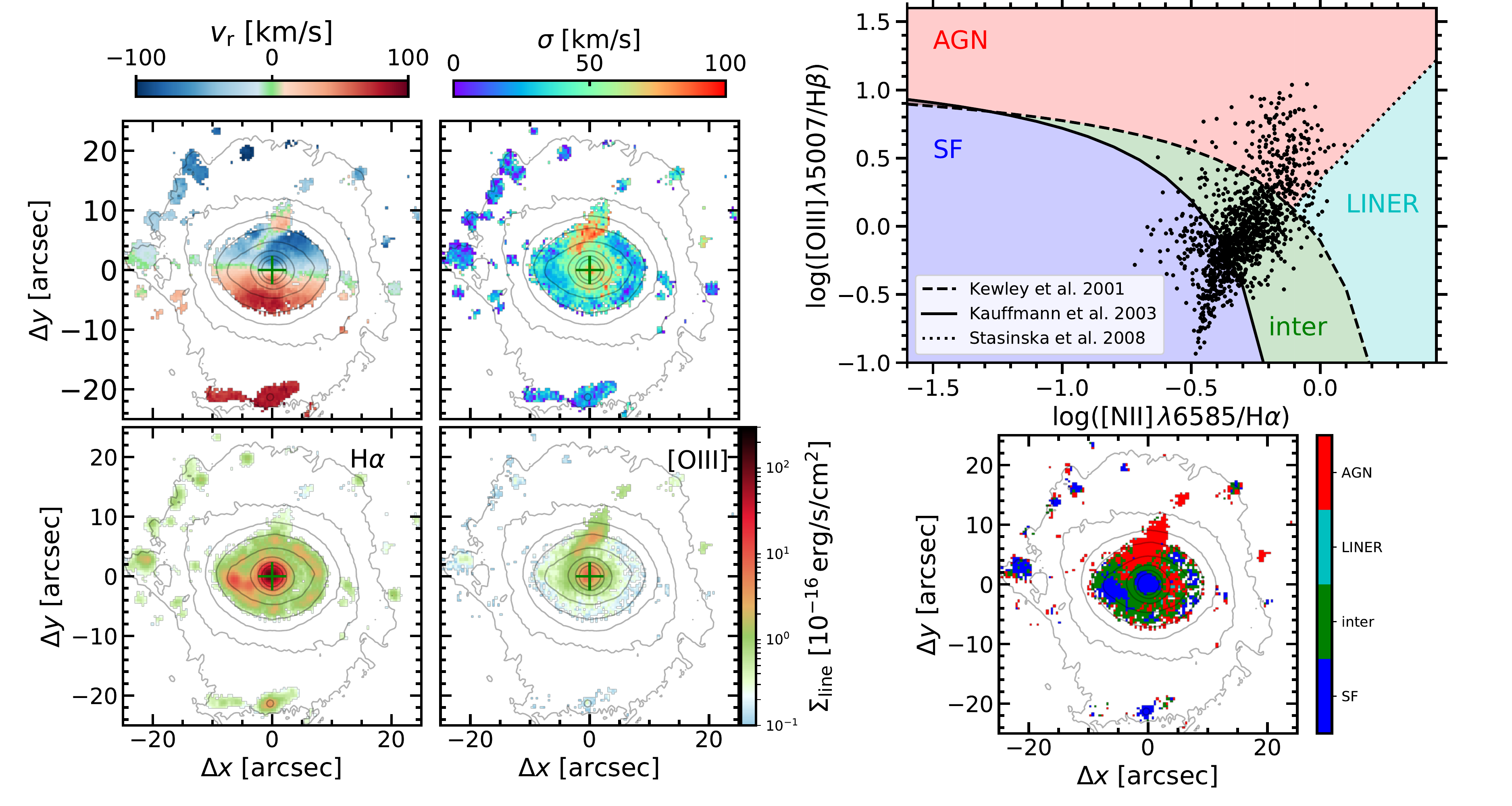}
 \caption{\textit{Left panels:} Ionized gas velocity and velocity dispersion maps (upper panels) for HE~0227$-$0913 together with the H$\alpha$ and [\ion{O}{iii}] surface brightness maps from the MUSE data after stellar continuum subtraction. Contours of the continuum emission are in logarithmic scaling are overlayed in gray. \textit{Right panels:} Emission-line diagnostic diagram for spaxels with a S/N$>3$ in each line (upper panel). Excitation map after separating line ratios, into star forming, AGN, intermediate and LINER-like ionization.}\label{fig:HE0227_line}
\vspace*{3.8mm}
 \includegraphics[width=\textwidth]{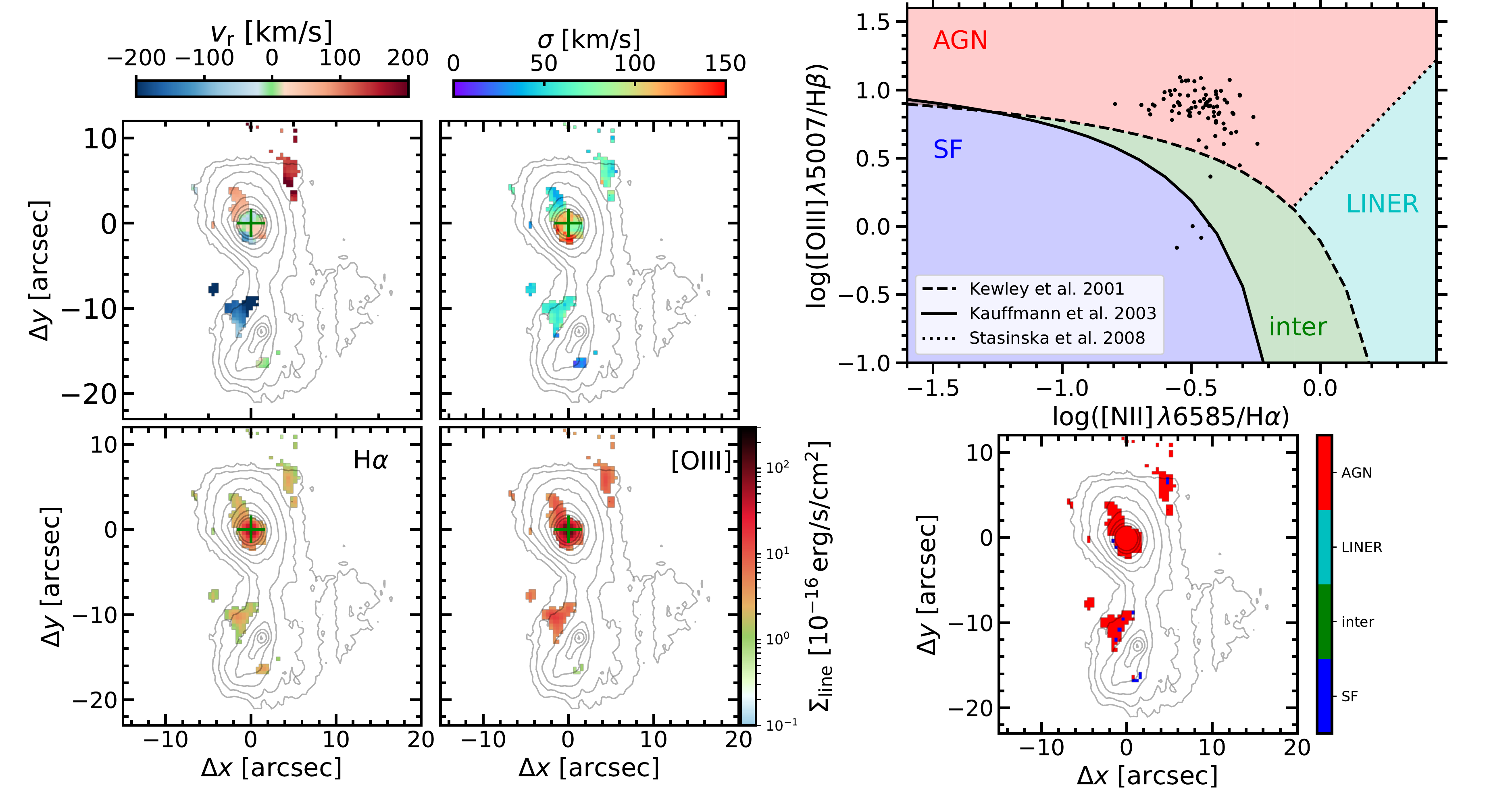}
 \caption{Same as Fig.~\ref{fig:HE0227_line}, for HE~0351$+$0240.}\label{fig:HE0351_line}
\end{figure*}

\subsection{Optical emission-line analysis}
Since HE~0227$-$0913 and HE~0351$+$0240 are both unobscured QSOs, we first have to subtract the point-like unresolved AGN emission before performing the emission line mapping across the two galaxies. Here, we use the AGN-host galaxy IFU deblending software QDeblend${}^\mathrm{3D}$ \citep{Husemann:2013,Husemann:2014}. The tool utilizes a simple nonparametric algorithm to firstly determine the on-axis PSF through intensity mapping of the broad H$\beta$ and H$\alpha$ lines, and secondly subtract the central AGN spectrum convolved by the PSF in an iterative manner to decontaminate the AGN spectrum from the host galaxy contributions at the center of the galaxy. The original spectrum within an aperture of 3\arcsec\ and the respective  AGN and host galaxy contribution are shown in Fig.~\ref{fig:decomp} for both galaxies. The narrow line components had a best-fit  velocity dispersion of $\sigma=$ 131 km/s for HE~0227--0913, and 65 km/s for HE~0351+0240.

As a next step, we fit the stellar continuum and emission lines across the QSO-subtracted data cube using PyParadise \citep{Husemann:2016a,Weaver:2018}. Since the S/N for both galaxies is not enough to model the continuum for individual spaxels over the entire galaxies, we first perform a Voronoi-binning scheme \citep{Cappellari:2003} to achieve a S/N of at least 20 per bin. The Voronoi-binned spectra are then fitted with a linear superposition of stellar spectra from the INDO-U.S. library \citep{Valdes:2004}, convolved with the MUSE mean spectral resolution of 2.4\AA, and a redshifted Gaussian kernel to match the kinematics. In a second step, we model the continuum in the same way for all individual spaxels but keeping the previously inferred kinematics fixed. After subtracting the best-fit continuum model, we fit all bright emission lines such as  H$\beta$, [\ion{O}{iii}] $\lambda\lambda$4960,5007, H$\alpha$, [\ion{N}{ii}] $\lambda\lambda6548,6583$ and [\ion{S}{ii}] $\lambda\lambda$6716,6730, with a Gaussian shape, again convolved with the instrumental resolution, where the redshift and intrinsic velocity dispersion are assumed to be the same for all lines. The [\ion{O}{iii}] and [\ion{N}{ii}] doublets are fixed in their line ratio according to the theoretical expectations \citep{Storey:2000}. Errors on all continuum and emission-line model parameters are estimated using a Monte-Carlo approach where we re-fit the spectra about 30 times after replacing each pixel within the normal distribution around the initial values with a width guided by the pipeline errors.

\begin{figure*}
   \centering          
   \includegraphics[width=\textwidth]{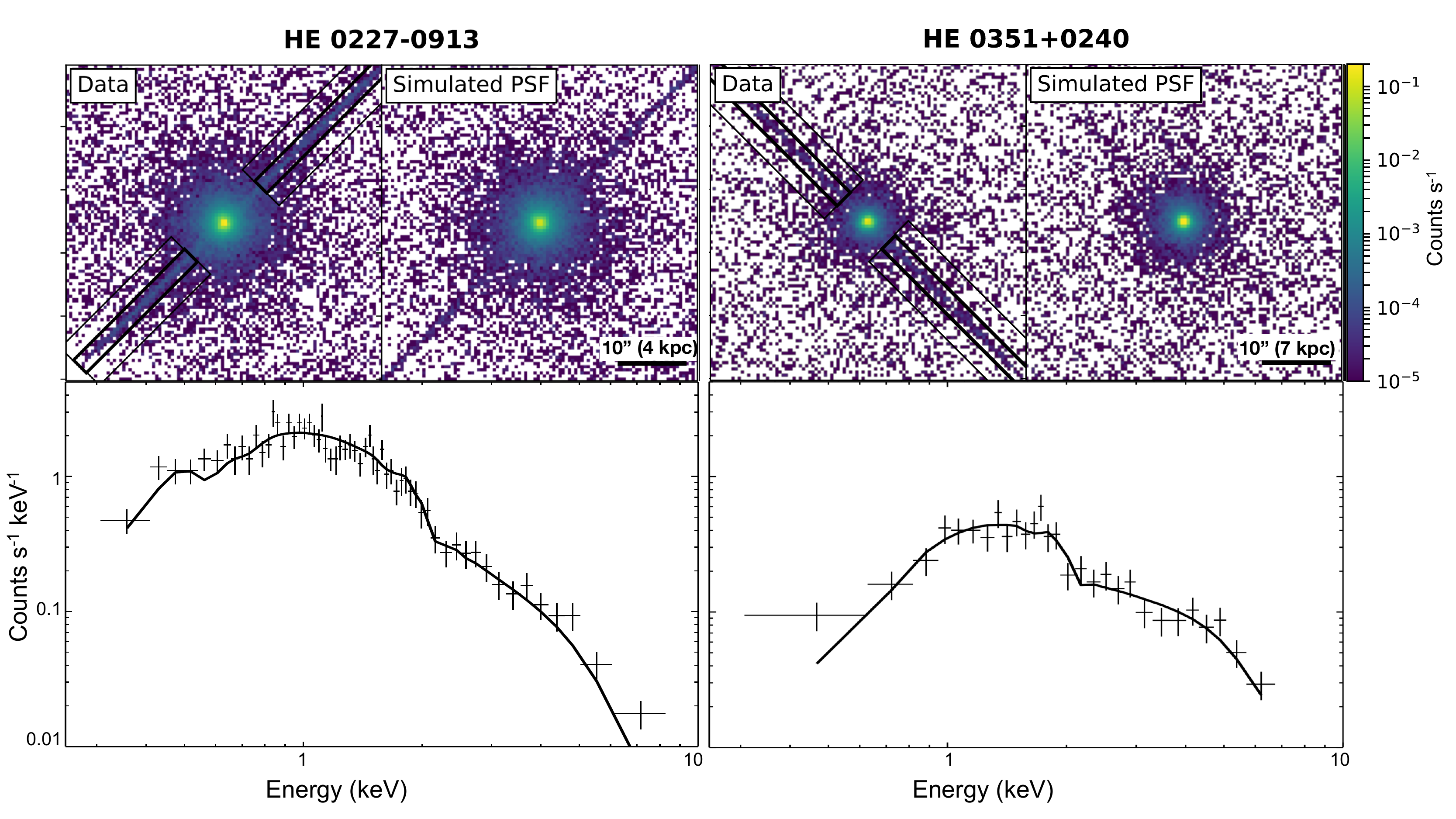}
      \caption{{\it Top}: Coadded {\it Chandra} images of HE~0227$-$0913 and HE~0351$+$0240 (with readout streaks) and the corresponding simulated PSFs. The color scale is in units of counts per second, shown by the right-hand colorbar.
      {\it Bottom}: Spectra of the co-added, background subtracted point sources taken from the readout streaks (shown in the thick black regions above, with adjacent background regions), binned to have a minimum significance of 5 per bin. The spectra were fit to a single power-law, with the \texttt{xspec} model: (\texttt{phabs*powlaw1d}). In each model we fixed Galactic absorption to the value reported in \texttt{XSPEC}, $3.26 \times 10^{20}$ cm$^{-2}$ for HE~0227$-$0913 and $1.28 \times 10^{21}$ cm$^{-2}$ for HE~0351$+$0240. The best-fit models (solid black lines) were used to simulate the PSFs (shown above).}
         \label{fig:xim}
   \end{figure*}
   
The fitting results for HE~0227$-$0913 and HE~0351$+$0240 are summarized in Fig.~\ref{fig:HE0227_line} and Fig.~\ref{fig:HE0351_line}, respectively, featuring the ionized gas velocity and velocity dispersion as well as the H$\alpha$ and [\ion{O}{iii}] surface brightness maps. Based on the classical emission-line diagnostic diagrams \citep[e.g.,][]{Baldwin:1981, Veilleux:1987}, we create an excitation map using demarcation lines in the [\ion{O}{iii}]/H$\beta$ vs. [\ion{N}{ii}]/H$\alpha$ to distinguish between star forming, AGN ionization, intermediate and LINER-like regions \citep{Kewley:2001,Kauffmann:2003,Stasinska:2008}. The resulting properties of both galaxies are quite different and we describe the main features in the following.

The host galaxy of HE~0227$-$0913 is a disk which is clearly supported by the ionized gas velocity field. This velocity field was modeled using the KinMS software\footnote{https://github.com/TimothyADavis/KinMS} \citep{Davis:2013} to obtain an inclination angle of $67.4\pm4.5^{\circ}$. Ongoing star formation dominates the gas excitation in a circumnuclear disk, with an inner star forming ring a few arcsec ($<1$ kpc) in diameter and an apparently outer star forming ring at $\sim$20\arcsec\ ($7$ kpc) distance that is likely related to the spiral arms. The presence of regions dominated by ionization from OB-stars is consistent with star formation within the ISM of the galaxy; it has an estimated \Htwo\ mass of $4\times 10^{8}~M_{\odot}$ \citep{Bertram:2007}, and an \HI\ mass of $2.6\times 10^{9}~M_{\odot}$ \citep{Konig:2009}. Another striking feature is a bright AGN-ionized arc-like feature of ionized gas about 5-15\arcsec\ (or 5 kpc) North of the nucleus, which is redshifted in velocity space with respect to the underlying disk motion but nearly at zero offset from the systemic velocity. The high velocity dispersion may not entirely be intrinsic to the AGN-ionized gas, but additionally broadened due to the superposition of the emission line kinematics of the gas disk. 

The interacting galaxy system of HE~0351$+$0240 does not show any strong signatures of ongoing star formation (there is no \COzero\ detection for this galaxy; \citealt{Bertram:2007}), as all of the ionized gas is ionized by the AGN and displays a filamentary structure with a complex kinematic pattern. The velocity dispersion is elevated close to the AGN nucleus, but not in the extended filamentary structures.

In the following we will test the hypothesis that the AGN-ionized gas filaments represent the interface of a slowly expanding hot gas bubble with the surrounding cool gas. Such hot gas bubbles are predicted by numerical simulation \citep{Weinberger:2017} and other theoretical considerations \citep{FQ:2012,Zubovas:2012} as a natural consequence of radiation-driven feedback scenarios.

\subsection{X-ray analysis}
We searched for extended X-ray emission by comparing our observations to the {\it Chandra} PSF, the accurate modeling of which requires the point-source spectrum. As Type 1 AGN, the nuclei of HE~0227$-$0913 and HE~0351$+$0240 are very bright in X-rays, with high enough fluxes for pileup to be significant in warping the spectrum. However, the shutterless ACIS detector on \emph{Chandra} cause photons from a bright source to be detected while data are being read out, forming a linear streak through the source in the image. We can therefore use photons in this readout streak to model the point source spectrum for each source, as only the unresolved bright AGN is represented in this region without the pileup issues. We used the CIAO tool `{\tt dmextract}' to extract the spectrum in each observation, and the tools `{\tt mkacisrmf}' and `{\tt mkarf}' to make the response files at the location of the central source. We extracted the background spectra in adjacent regions next to the readout streaks, and we corrected the exposure time of the readout streak in order to model the correct flux. Lastly, we stacked the individual observations for each galaxy together using the CIAO script `{\tt combine\_spectra}' and used \texttt{SHERPA} to model the 0.3--7\,keV background-subtracted spectrum. 
   
AGN X-ray spectra are typically well modeled by a power-law, parametrized only by the photon index, normalization, and absorption. Fixing the absorption due to the galaxy and binning the spectrum to have a minimum SNR of 5 per bin, both galaxies were well fit by this model (`\texttt{phabs*powlaw1d}'). The fits are shown in Fig.~\ref{fig:xim} (lower panels) and the parameters of each model are listed in Table 2.
NLS1 galaxies like HE~0227$-$0913 are known to have soft X-ray excess above the powerlaw continuum, and indeed, soft excess in HE~0227$-$0913 has been previously found and studied with higher-resolution instruments \citep{Dewangan:2007,Mallick:2018}. These studies require multiple components to adequately fit the full spectrum. However, the improvement of our fit to the streak spectrum by including an additional component brings the reduced $\chi^{2}$ below 1, and the resulting PSF does not change our conclusions. We therefore use the 1-component powerlaw fit for the AGN in both galaxies. 

   \begin{figure*}
   \centering
	\includegraphics[width=\textwidth]{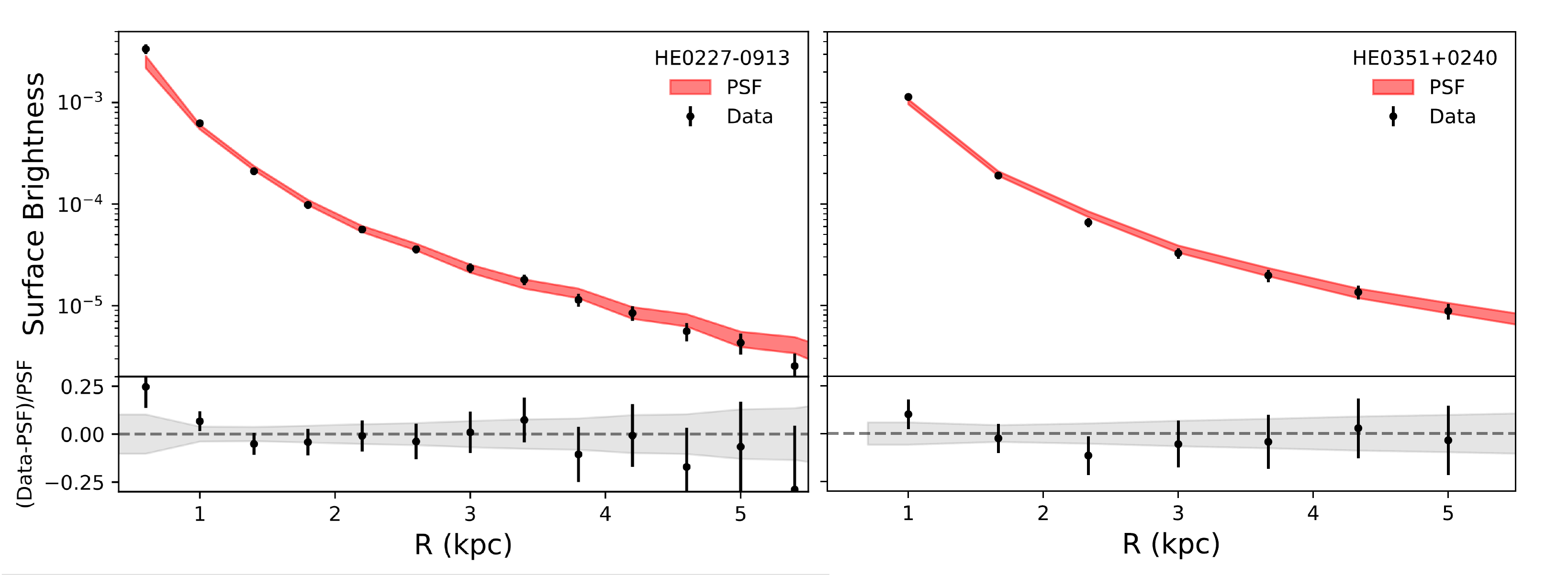}
      \caption{Radial profiles of the background-subtracted data (black cirlces) and of the simulated PSF (red lines) in bins of 1\arcsec, shown with 95\% level errors. Both 0.5--7\,keV data and PSF were used, excluding the readout streak regions. Bottom panels show the fractional residual errors. No clear evidence for extended emission is detected for  either galaxy.}
         \label{fig:rprof}
   \end{figure*}
   
\begin{table}

\label{table:2}    
\centering                          
\caption{X-ray spectral parameters for power-law model.}
\begin{tabular}{c c c}        
\hline   \hline             
Galaxy & Parameter & Value\\    
\hline                    
     		 &  &      \\
   HE 0227-0913 & $\Gamma_{PL}$ & $ 2.38 \pm 0.06$  \\      
     		& norm$_{PL}$ & $(6.78 \pm 0.21) \times 10^{-3}$   \\
            &   $\chi^{2}$/dof & 0.98\\
            & $f_{0.3-10\,\mathrm{keV}}$ & $2.12\times 10^{-11}$\\
            & $L_{2-10\,\mathrm{keV}}$ &  $9.2\times 10^{42}$    \\
      &  &      \\
   HE 0351+0240 & $\Gamma_{PL}$ & $1.38\pm 0.08$\\      
     & norm$_{PL}$ & $(1.52 \pm 0.10) \times 10^{-3}$    \\
      &   $\chi^{2}$/dof & 1.06 \\
      & $f_{0.3-10\,\mathrm{keV}}$ & $1.19\times 10^{-11}$\\
       & $L_{2-10\,\mathrm{keV}}$ &  $3.6\times 10^{43}$    \\

\hline
\end{tabular}   
\tablefoot{
The 0.3--10\,keV (absorbed) flux is in units of erg cm$^{-2}$ s$^{-1}$. The luminosities are in units of erg s$^{-1}$ and are corrected for Galactic absorption. The model fits were performed on the binned data using \texttt{Sherpa}.
}
\end{table}

To simulate the PSF we used the \texttt{ChaRT} online ray tracing tool \citep{Carter:2003}, where we input the modeled source spectrum, and then used the \texttt{MARX} software to project the rays onto the ACIS-S detector. We corrected for SIM drift and offset, and set the `AspectBlur' parameter to 0.19" to avoid pixelization effects in the pseudo-event file (the standard value for ACIS-S). The `{\tt marxpileup}' program was used to simulate pileup, where the `FrameTime' parameter was set to 0.7 seconds, as it is reduced when using 1/8 of the chip. The coadded data are shown in Figure \ref{fig:xim}, along with the simulated PSF, shown in a log color scale.

We used the CIAO tool \texttt{srcflux} to obtain the X-ray fluxes of each source using the resulting PSF models. We calculated the 0.3--10\,keV flux to be $2.12\times 10^{-11}$ erg cm$^{-2}$ s$^{-1}$ for HE~0227-0913 (consistent with \citealt{Dewangan:2007}) and $1.91\times 10^{-11}$ erg cm$^{-2}$ s$^{-1}$ for HE 0351+0240. This corresponds to luminosities of $L_{0.3-10\,\mathrm{keV}}=10^{43.1}$ erg s$^{-1}$ and $L_{0.3-10\,\mathrm{keV}}=10^{43.6}$ erg s$^{-1}$, respectively. Assuming a 2--10\,keV bolometric correction of 20 \citep{Lusso:2012}, we estimate the Eddington ratios ($\lambda_{\rm{Edd}} \equiv L_{bol}/L_{Edd}$) to be $\sim 1$ for HE 0227--0913 and $\sim 0.5$ for HE 0351+0240.

We compared the radial surface brightness profiles of the data to the PSF images to look for evidence of extended emission. We used the CIAO tool `{\tt dmextract}' to extract the events in 20 concentric annuli centered on each source from 1\arcsec\ to 20\arcsec\ (excluding the region of the readout streak). To estimate systematic errors from the incorrect pileup simulation, we estimated the fraction of counts affected by pileup via the \texttt{'pileup\_map'} tool, which decreases as a function of radius, and added this error to the Poisson error. The surface brightness profiles of the PSF and source for both galaxies are shown in Figure \ref{fig:rprof}, along with the fractional residual errors. We do not find any significant evidence for excess emission above the PSF, except in the first bin for HE~0227$-$0913; we find a $\sim 2-\sigma$ excess at $R<1$ kpc, however the systematic uncertainty due to pileup in the central region makes this detection very tentative. On scales of the ionized gas structures ($\sim 3-5$ kpc) no excess emission is found, even after dividing the data into soft and hard bins.

\begin{figure}
    \resizebox{\hsize}{!}{\includegraphics[width=\textwidth]{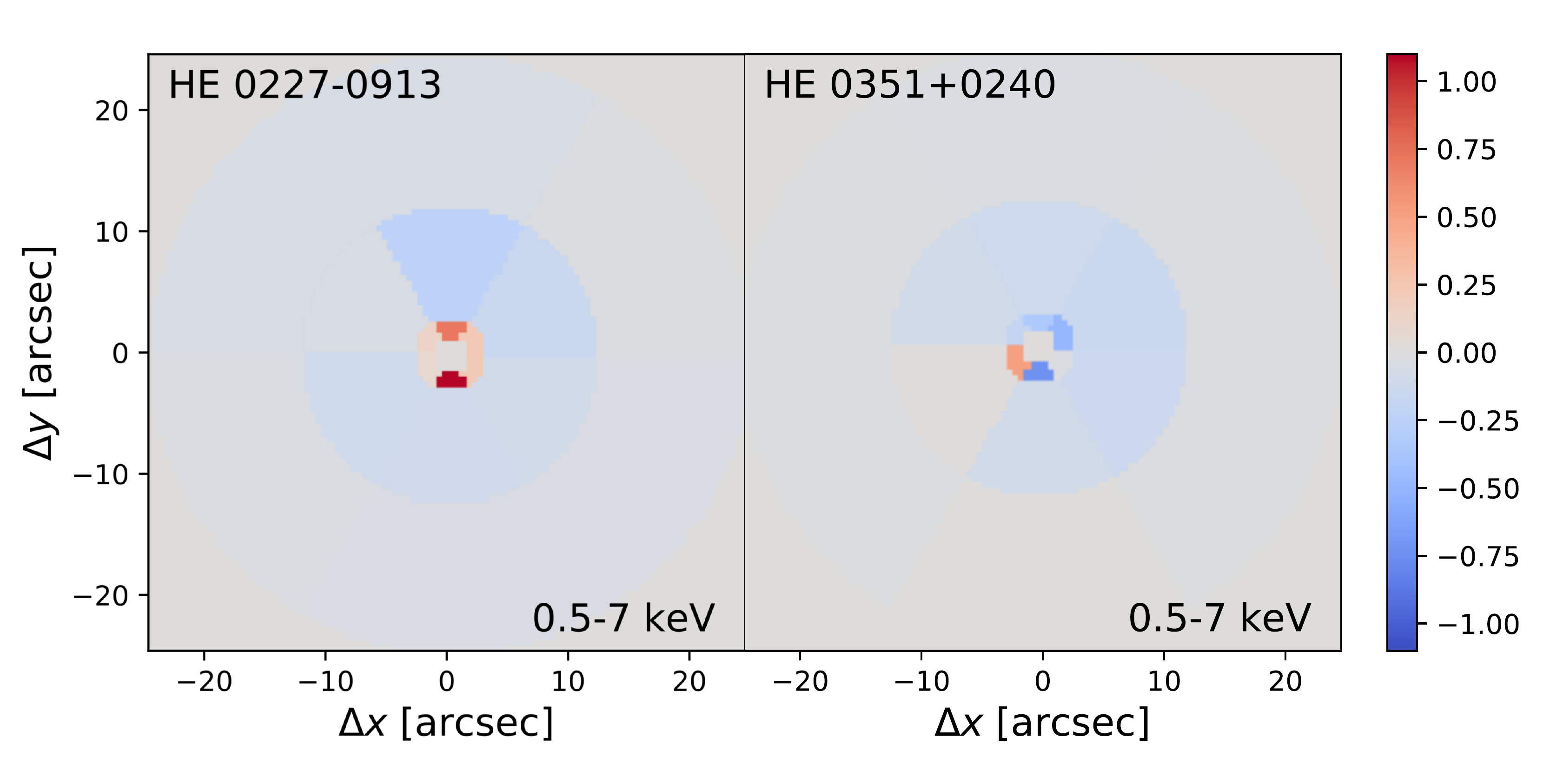}}
      \vspace*{-5mm}
      \caption{Difference between the data and PSF of the binned 2D images, divided by error for HE~0227$-$0913 (left) and HE~0351$+$0240 (right). No significant emission is detected beyond a 1.5$\sigma$ excess in the first radial bin of HE~0227$-$0913 (1\farcs5-3\arcsec).}
         \label{fig:az}
   \end{figure}

We further compared the flux in radial {\it and} azimuthal bins, to find any significant angular deviations from the PSF correlated with the ionized gas structures. Figure \ref{fig:az} shows the residual fluctuation maps (data minus PSF divided by the rms) in three radial bins and six azimuthal bins (the increasing size of the radial bins were to have comparable signal-to-noise in each bin). This shows there are no significant deviations from the modeled PSF when including pileup modeling errors.

\section{Discussion}

\begin{figure*}
   \centering
      \includegraphics[width=\textwidth]{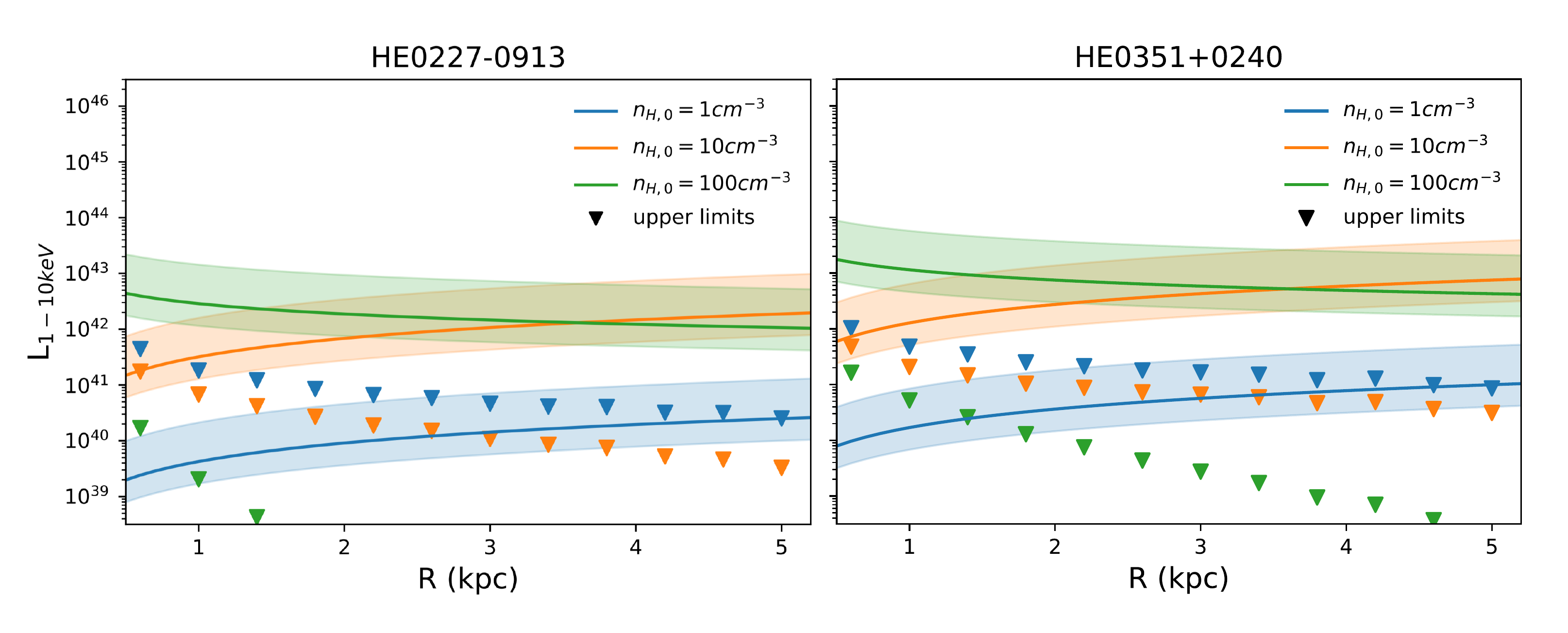}
      \caption{Prediction for X-ray bubble luminosity as a function of shock-front radius for three values of the interstellar gas density at R=100 pc ($n_{H,0}$) from the analytic model in \cite{Nims:2015}, assuming $n_{H}\propto R^{-1}$ (solid lines), compared to the upper limits from our observations (solid triangles). Different upper limits are shown because the predicted wind spectrum depends on density. The errors on the prediction come from bolometric correction uncertainty ($\kappa_{2-10\,\mathrm{keV}}=20^{+80}_{-12}$) when scaling to the bolometric luminosities of HE~0227$-$0913 and HE~0351$+$0240. Except for the lowest densities, our upper limits fall below the prediction.}
         \label{fig:nims}
   \end{figure*}
   
\subsection{Comparison to AGN wind theory}

\citet{Nims:2015} investigated the observational signatures of AGN outflows, and predicted the resulting X-ray spectrum and luminosity of shocked ambient material due to an AGN-driven wind out to 2 kpc. Assuming the kinetic outflow luminosity is $5\%$ of the AGN luminosity, they predict 1--10\,keV luminosities for various forward shock radii, which depends on the parameters of the ambient density profile (assuming a spherically symmetric profile $n_{H}=n_{H,0}\times (R_S/R_0)^{-1}$; \citealt{FQ:2012}) and the bolometric luminosity of the AGN. Using fiducial values motivated by observations of nearby quasars and ULIRGs \citep{Feruglio:2010,Cicone:2014}, they showed that on kpc scales, the X-ray luminosity is dominated by free-free emission.

We compared their predictions to our results in the following way: we calculated the $3\sigma$ upper limits of our simulated PSFs and used {\tt PIMMS} to convert the surface brightness to 1--10\,keV flux, by assuming a thermal bremsstrahlung spectrum with a temperature in the analytic form for the shocked ambient medium in \citeauthor{Nims:2015}, for three density values. We then scaled the Nims prediction for each density to the bolometric luminosities of the AGN, assuming a bolometric correction $\kappa_{2-10\,\mathrm{keV}}=20^{+80}_{-12}$ \citep{Lusso:2012}, and extrapolated to 5 kpc. We note that this bolometric correction is conservative for Mrk 1044 because its black hole is super-Eddington; previous work looking at its SED has estimated it to have a bolometric luminosity an order of magnitude larger ($\sim 10^{46}$ erg s$^{-1}$; \citealt{Castello:2016}), which would predict a more luminous wind. The comparison plots are shown in Figure \ref{fig:nims}. For density values (at a radius of 100 pc) of $10-100$ cm$^{-3}$, our upper limits are well under the prediction. However for low densities $\sim 1$ cm$^{-3}$ the prediction is much fainter, $L_{1-10\,\mathrm{keV}}=10^{39}-10^{41}$ erg s$^{-1}$, at or below our upper limits on the scales of the ionized gas. This suggests that either (1) the ambient medium density is low ($\leq 1$ cm$^{-3}$) such that the extended emission is undetectable, or (2) the coupling efficiency between the AGN luminosity and the kinetic luminosity is less than predicted by \citealt{Nims:2015} ($L_{\rm{kin}} < 0.05L_{\rm{AGN}}$). 

For the first case, it is possible that the fiducial values for the ambient density, inferred from winds in nearby ULIRGs\footnote{The fiducial density values are much lower than the measured mean molecular gas densities in the nuclear regions of ULIRGs \citep{Downes:1998} since winds are thought to propagate along the paths of least resistance \citep{FQ:2012}.} \citep{Cicone:2014,Feruglio:2010}, are very different from the host galaxies in this study. If the environmental properties of these ULIRGs turn out not to be representative of typical Type 1 AGN host galaxies, then our nondetection may suggest lower densities of at least an order of magnitude. For the second case, there have been a wide range of estimates of the coupling efficiency derived from observations, anywhere between 0.001\% and 10\% \citep{Harrison:2018}. 5\% has been a standard value in the literature, obtained first from hydrodynamic simulations as the value needed to recover the local $M_{\rm{BH}}-\sigma$ relation \citep{DiMatteo:2005}. This value also comes up naturally from theoretical considerations of momentum-driven nuclear outflows with speeds of $\sim 0.1 c$ \citep{King:2003,Harrison:2018}, and indeed, this may be appropriate for Mrk 1044, as \cite{Mallick:2018} tentatively detected an ultra fast outflow (with a speed of $\sim 0.1 c$) from a deep {\it XMM-Newton} observation of its nucleus. However, several recent observations and simulations of AGN suggest much smaller coupling efficiencies ($\sim 0.5\%$; \citealt{Husemann:2016c,Veilleux:2017,Richings:2018}), which would be consistent with our results.

There have been a few previous cases of extended X-ray emission found in AGN possibly from AGN superwinds. \cite{Greene:2014} found extended X-ray emission in SDSS J1356+1026, a ULIRG galaxy in an ongoing merger with a 20 kpc bipolar outflow. The non-nuclear emission was estimated to be $L_{0.3-8\,\mathrm{keV}}=7\times 10^{41}$ erg s$^{-1}$, with densities of $\sim 2.4$ cm$^{-3}$ at the kpc scale and an AGN luminosity of $L_{bol}=10^{46}$ erg s$^{-1}$. In another example, the teacup galaxy ($L_{bol}=10^{45}-10^{46}$ erg s$^{-1}$) has X-ray emission ($L_{2-10\,\mathrm{keV}} \sim 10^{44}$ erg s$^{-1}$), cospatial with its ionized, 10 kpc bubble \citep{Lansbury:2018}. The spectrum is consistent with a two-temperature $\sim 0.4$ keV and $\geq 2.7$ keV thermal gas, with $L_{1-10\,\mathrm{keV}}\sim 10^{41}$ erg s$^{-1}$. For both galaxies, we would have detected their diffuse hot gas were it present in our objects. However, the X-ray luminosities fall below the predicted values for their respective AGN luminosities. Mrk 231 ($L_{bol}=10^{46.2}$ erg s$^{-1}$) also has diffuse X-ray emission fainter than predicted, with $L_{2-10\,\mathrm{keV}}\sim 10^{41}$ erg s$^{-1}$ within the first few kpc \citep{Veilleux:2014}.

\subsection{Other sources of extended X-ray emission}
Extended X-ray emission can also come from hot halo gas, typically found in galaxy groups and clusters.
The stellar masses of the two galaxies studied here are $\log (M_\star/M_{\odot}h^{-1})\sim 10.1$ and $10.5$, estimated based on the $g-i$ color and $i$ band absolute magnitude from the QSO-subtracted broadband MUSE images following the prescription of \citet{Taylor:2011}. This corresponds to halo masses of $\log (M_h/M_{\odot}h^{-1})\sim 11.6$ and 12.0 assuming the stellar mass to halo mass relation of \cite{Behroozi:2010}. At these mass scales, the expected total X-ray luminosity from extended emission is $\lesssim 10^{39}$ erg s$^{-1}$ \citep{Anderson:2015}, which is well below the superbubble predictions on the largest scales, below our detection limit.

Star formation in galaxies is another source of X-rays, produced by the evolved X-ray binaries as well as supernovae. While this is a negligible contribution for HE~0351+0240 due to the lack of star forming regions, the star formation rate in the nuclear region of Mrk 1044 is $\sim 1~M_{\odot}/yr$ within the inner 3\arcsec\ (inner kpc) based on the extinction-corrected H$\alpha$ luminosity following \citet{Kennicutt:1998}. This SFR corresponds to an X-ray luminosity of $L_{0.5-8\,\mathrm{keV}}\sim 4\times 10^{39}$ erg s$^{-1}$ \citep{Mineo:2014}. While this is below our upper limit (and the superbubble predictions for the mid and high density values), it could be contributing to the tentative detection we find in the first radial bin.

\subsection{The origin of highly ionized gas arcs}
The two AGN host galaxies were selected from CARS for our {\it Chandra} pilot observations because of the prominent arc-like structures of highly-ionized gas on kpc scales from the AGN. Based on the deep {\it Chandra} observations we argue that the arcs are most likely not related to the interplay between the hot medium of an AGN-driven outflow bubble and the cold ambient medium for these two galaxies. While hot gas AGN-driven bubbles may be important for other AGN hosts, it does not explain the existence of the prominent ionized gas arcs in HE 0227-0913 and HE 0351+0240. So what are the alternative explanations for their appearance and shape in these two studied cases?

Photoionization of gas by the AGN on several kpc scales is usually referred to as the extended narrow-line region (ENLR) or the extended emission-line region (EELR), which appears to be a common feature around luminous radio-loud \citep[e.g.,][]{Stockton:1987,Villar-Martin:1997,Fu:2009} and radio-quiet AGN \citep[e.g.,][]{Bennert:2002,Husemann:2013,Liu:2013,Hainline:2013,Harrison:2014,Sun:2017}. The size, geometry and shape of the ENLR potentially depends on several factors, such as the AGN luminosity \citep[e.g.,][]{Husemann:2014,Sun:2017,Sun:2018}, the opening angle and orientation of the AGN ionization cones \citep[e.g.,][]{Bennert:2006c}, the gas density distribution intercepting the ionizing radiation field of the AGN \citep[e.g.,][]{Dempsey:2018}, or other intrinsic properties of the central AGN engine \citep{Husemann:2008,Keel:2012}. The intrinsic AGN luminosity is comparable in both galaxies studied and corresponds to expected ENLR sizes of 5-10 kpc depending on different ENLR size definitions and samples \citep{Husemann:2014, Sun:2017}. Since the morphologies of the host galaxies are rather different, we expect that the relation between the geometry of the AGN radiation field and the distribution of the gas plays the dominant role in shaping the appearance of the [\ion{O}{iii}] arcs, as we discuss below.

HE~0227$-$0913 shows a disk-dominated morphology, which is also evident in the rotation-dominated velocity field of the star-forming regions (see Fig.~\ref{fig:HE0227_line}). The velocity of the [\ion{O}{iii}] arc clearly does not follow that rotational motion and is close to the systemic velocity of the galaxy. One possibility is that the gas was tidally stripped from a minor companion galaxy with a close orbit around HE~0227$-$0913. However, there is no signature of such a companion galaxy in the continuum light that could be spatially associated with the ionized gas arc. Unless the stellar component was tidally stripped as well, resulting in an undetectable low surface brightness companion, this scenario is unlikely. Furthermore, ionization by shocks can be excluded since the high [\ion{O}{iii}]/H$\beta\sim10$ ratio can only be produced by fast shocks ($>500\,\mathrm{km\,s}^{-1}$) and their precursors \citep{Allen:2008}, which are not supported by the quiescent kinematics of the gas. Instead, we noticed that the highest surface brightness regions of the [\ion{O}{iii}] arc appear at a similar radius to the star forming ring visible in the stellar continuum and H$\alpha$ emission. Therefore, an alternative possibility is that the high star formation surface density in the ring has caused a gas outflow perpendicular to the galaxy disk, as predicted in the Galactic fountain picture \citep[e.g.,][]{Shapiro:1976,Fraternali:2006}. Recent simulations show that the warm gas will likely fall back to the disk \citep[e.g.,][]{Kim:2018}. The illumination of clouds ejected from the galaxy and falling back seems a plausible scenario for HE~0227$-$0913, as the AGN radiation field can only significantly intercept gas above the disk at radii of a few kpc. 

On the other hand, HE~0351$+$0240 is clearly an ongoing major merger system. During this process, the gas in the two parent galaxies have likely been significantly redistributed due to their tidal interactions. Gas shells and streams can be easily created on large scales, depending on the orbital configuration and angular momenta of interacting galaxies \citep[e.g.,][]{Arp:1966,Hibbard:1996}, and predicted by numerical simulation of mergers \citep[e.g.,][]{Toomre:1972,Mihos:1996}. If the AGN radiation field is intercepting those streams, they become visible as highly ionized gas as happens, for example, in the major merger NGC~7252 \citep{Schweizer:2013,Weaver:2018}. Since ionization by shocks can be excluded with the same argument as for HE~0227$-$0913, we argue that the [\ion{O}{iii}] arcs of HE~0351$+$0240 are likely produced as the intersection of the large-scale distribution of tidal gas streams and the AGN radiation field. Assuming a static ionization cone, the opening angle would have to be $\gtrsim 80^{\circ}$; however the central engine likely changes orientation with respect to the larger gas environment due to the disturbed kinematics of the merging system.

\section{Summary and conclusions}

In this study, we united MUSE IFU data with deep {\it Chandra} observations to search for quasar-driven superbubbles in two nearby AGN host galaxies from the CARS survey. Selected based on their quiescent kpc-scale  emission arcs, both galaxies were promising candidates for catching the hot-wind mode of radiative AGN feedback in action. 
However, we found no evidence for extended, hot diffuse gas associated with a super wind. 
Our upper X-ray luminosity limits of $\sim 10^{40-41}$ erg s$^{-1}$ on kpc scales is well below the expectation ($\sim 10^{42}$ erg s$^{-1}$) based on standard theory \citep{Nims:2015}.
From this nondetection, we conclude the following:
\begin{itemize}
	\item{Based on theoretical predictions, either the coupling efficiency between the AGN and ambient medium must be small ($<5\%$), or the density of the medium through which the wind travels is low ($n_{H}\lesssim 1$ cm$^{-3}$  at a distance of 100 pc).}
	\item{Photoionization from the AGN of clouds that were previously ejected from the galaxies is likely the origin of the ionized gas structures observed in each AGN.}
	\item{Hot winds are either not ubiquitous among moderate-luminosity AGN, or are not connected to the observed ionized kpc-scale structures that are common in such objects.}
\end{itemize}

While AGN-driven winds may be present for the highest-luminosity AGN, their extent and their relation to the ionized gas component are still not well constrained. The physics of AGN outflows remains murky, and only multiphase ISM observations can reveal the full picture \citep{Cicone:2018,Husemann:2018b}. For this purpose, future investigations with AGN from CARS are planned, utilizing spatially-resolved data across the full electromagnetic spectrum.

\begin{acknowledgements}
Support for this work was provided by the National Aeronautics and Space Administration through \textit{Chandra} Award Number GO6-17093X issued by the \textit{Chandra} X-ray Center, which is operated by the Smithsonian Astrophysical Observatory for and on behalf of the National Aeronautics Space Administration under contract NAS8-03060. 
GRT acknowledges support from NASA through
\textit{Chandra} Award Number GO7-8128X as well as Einstein Postdoctoral Fellowship
Award Number PF-150128, issued by the \textit{Chandra} X-ray Observatory
Center, which is operated by the Smithsonian Astrophysical
Observatory for and on behalf of NASA under contract NAS8-03060.
MK acknowledges support by DFG grant KR 3338/3-1, and TAD acknowledges support from a Science and Technology Facilities Council Ernest Rutherford Fellowship. MPT acknowledges support by the Spanish MINECO through grants AYA2012–38491–C02–02 and AYA2015–63939–C2–1–P, co-funded with FEDER funds. 

\end{acknowledgements}

%-------------------------------------------------------------------

\bibliography{references}
\bibliographystyle{aa}
\end{document}